\documentclass[journal]{IEEEtran}

\usepackage{amsmath,amssymb}
\usepackage{graphicx}
\graphicspath{{fig/}}
\usepackage{array}
\usepackage[caption=false,font=footnotesize]{subfig}

\usepackage{booktabs}
\usepackage{tabularx}
\newcolumntype{Y}[1]{>{\raggedright\arraybackslash\hsize=#1\hsize}X}
\newcolumntype{R}[1]{>{\raggedright\arraybackslash}p{#1}}

\usepackage{url}
\usepackage[hidelinks]{hyperref}

\usepackage{cite}

\usepackage[american]{babel}

\usepackage{tikz}
\usetikzlibrary{arrows.meta,positioning,calc}
\tikzset{
  block/.style={draw=black!70, line width=.45pt, rounded corners=.6pt,
    align=center, font=\footnotesize, minimum height=8mm,
    inner xsep=2mm, inner ysep=1.5mm},
  input/.style={block, fill=black!4},
  process/.style={block, fill=blue!5},
  output/.style={block, fill=black!10},
  arrow/.style={-{Latex[length=1.7mm]}, line width=.5pt},
  note/.style={font=\scriptsize, align=center}
}

\begin{document}

\title{Nonlinear Interference Estimation in Coherent Optical Systems:
Principles and Applications}

\author{Dario~Pilori,~\IEEEmembership{Senior~Member,~IEEE}%
\thanks{Dario Pilori is with the Department of Electronics and Telecommunications (DET), Politecnico di Torino,
Torino, Italy. E-mail: dario.pilori@polito.it.}%
\thanks{Manuscript received XXX YYYY 2026.}}

\markboth{Journal of Lightwave Technology,~Vol.~XX, No.~Y, ZZZZZZ~2026}%
{Pilori: Nonlinear Interference Estimation}
\IEEEspecialpapernotice{(Invited Paper)}

\maketitle

\begin{abstract}
Accurate estimation of nonlinear interference (NLI) is important for monitoring and optimizing coherent 
optical networks. This invited paper reviews the main techniques proposed for receiver-side NLI estimation, 
with particular focus on longitudinal power monitoring (LPM). We discuss the principles, advantages and 
limitations of LPM-based NLI estimation, including recent experimental demonstrations. We then
experimentally investigate the impact of the spatial step size and the number of averaged constellation blocks on estimation accuracy. 
The results show that accurate NLI estimation does not necessarily require a highly accurate power profile, with 
limited improvements obtained by increasing the number of averaged blocks. Finally, we discuss the main 
practical limitations and applications of NLI estimation, with particular attention to end-to-end and 
span-wise optical power optimization.
\end{abstract}


\IEEEpeerreviewmaketitle

\section{Introduction}
\IEEEPARstart{I}{n} modern optical networks, accurate
monitoring is essential for their operation \cite{castoldi:jocn2025}.
In particular, the availability of telemetry information from network equipment allows
network operators to detect issues early, optimize transmission performance, reduce
operating margins and build accurate digital twins of the network 
\cite{zhuge:jocn2023,wang:jlt2024,nishizawa:jocn2026}.
Among monitoring techniques, those based on the receiver's digital
signal processing (DSP) \cite{faruk:jlt2024} are particularly interesting, 
since they can provide useful information without requiring additional hardware.
This approach is especially attractive in open optical networks, where
transceivers and optical line systems can be operated by different entities
\cite{kaeval:jocn2025}.
In fact, modern coherent transceivers already estimate several parameters \cite{hauske:jlt2009,dong:jlt2016},
such as chromatic dispersion (CD) \cite{sui:jlt2013}, polarization-dependent
loss (PDL) \cite{geyer:ptl2008} and the overall signal-to-noise ratio (SNR).
However, other important parameters remain significantly more challenging
to estimate.

One of them is the power of the nonlinear interference (NLI) \cite{dar:opex2013} affecting the channel,
generated by the Kerr effect during propagation in the optical fiber.
While the overall SNR is normally estimated by coherent receivers, as it is needed to compute the log-likelihood 
ratios (LLRs) for soft decoding, separating NLI from other noise sources is notoriously hard 
\cite{pilori:sppcom2025}. 
Nevertheless, the knowledge of NLI power would enable straightforward optimization of the channel launch power
\cite{poggiolini:jlt2014} and thus channel performance,
without relying on additional knowledge of network parameters. 
Several methods have been proposed to estimate NLI in coherent receivers
\cite{dong:opex2012,vaquerocaballero:jocn2018,lippiatt:ofc2020,shiner:ofc2020,dirosa:opex2020,lu:opex2020,
cho:opex22,vaquerocaballero:jlt2022,andrenacci:jocn2024,alnahhal:jlt2024,boertjes:jlt2024,bajaj:ondm2024,
guo:ecoc2024,girardjollet:ofc2025,bajaj:ecoc2026,girardjollet:ecoc2026}.
Usually, they exploit the statistical properties or the spectral characteristics of NLI, for instance by analyzing the 
received constellation or the information extracted from the receiver DSP. 
However, all these approaches have some limitations, such as the need for calibration, prior knowledge of the 
transmission link, or modifications to the transmitter. Therefore, obtaining reliable NLI estimates by passively 
monitoring the receiver DSP, with limited knowledge of the link, remains an important research challenge.

Recently, longitudinal power monitoring (LPM) has emerged as a viable alternative technique for NLI estimation.
LPM was first proposed \cite{tanimura:ecoc2019pdp} as a technique to estimate the
optical power evolution (power profile) of a channel, using only information
already available at the receiver \cite{sasai:jlt2022}.
However, to compute the power profile, LPM must also estimate the NLI at the receiver. 
Therefore, linear least-squares (LLS) LPM can, in principle, also be used for NLI estimation. 
This application was first theorized in \cite{sasai:jlt2022},
demonstrated in simulations by re-propagating the estimated profile \cite{kim:ofc2024}
and directly from LLS-LPM experiments \cite{andrenacci:jlt2025}.
Finally, it was implemented in a coherent DSP chip, as demonstrated in \cite{sasai:ecoc2026invited}.

In this invited paper, 
extending the invited presentation at ECOC 2026 \cite{pilori:ecoc2026invited}, 
we review the main principles and recent advances in NLI estimation for coherent optical systems. 
We first introduce the fundamental problem of NLI estimation 
and compare the main techniques proposed in the literature, 
highlighting their advantages and limitations. Then, we focus on LPM-based NLI estimation, reviewing its main principles 
and recent experimental demonstrations. 
In addition, we present a new experimental analysis on the impact of the spatial step size and of the number of averaged constellation blocks on NLI estimation accuracy.
Finally, we discuss the applications of NLI estimation to optical network 
optimization, with particular attention to launch power optimization 
and the prospects of span-wise estimation.

\section{NLI Estimation}\label{sec:nliest}
\subsection{Noise Segregation}
It has been widely demonstrated that an optically-amplified coherent optical
link can be accurately modeled as an additive white Gaussian noise (AWGN) channel
\cite{poggiolini:jlt2014}. The noise comes from three main sources, which
are statistically independent of each other. Consequently, the total
SNR, measured on the constellation after DSP, can be computed as
\begin{equation}\label{eq:snr}
    \mathrm{SNR} = (\mathrm{SNR}_{\mathrm{TRX}}^{-1} + \mathrm{OSNR}^{-1} + \mathrm{SNR}_{\mathrm{NL}}^{-1})^{-1} \,.
\end{equation}
$\mathrm{SNR}_{\mathrm{TRX}}$ accounts for transceiver noise, $\mathrm{OSNR}$ for 
amplified spontaneous emission (ASE) noise and $\mathrm{SNR}_{\mathrm{NL}}$ for Kerr-induced NLI; 
$\mathrm{OSNR}$ is computed on a reference bandwidth equal to the symbol rate.
This expression is particularly important for network operators, since the
OSNR scales with the launch power of the channel $P_\textup{ch}$, while
$\mathrm{SNR}_{\mathrm{NL}}$ scales with the inverse of its square $P^{-2}_\textup{ch}$.
The transceiver SNR, instead, is largely independent of this value, and 
it depends on the transceiver technology and its input power. Therefore, it will
be left out of this analysis.
This simple approximation has a direct consequence: the optimal
launch power $P_\textup{opt}$ is obtained when the
$\mathrm{SNR_{NL}}$ is 3~dB higher than the OSNR
\begin{equation}\label{eq:distFromOpt}
    P_\mathrm{opt} - P_\mathrm{ch} = \frac{1}{3}\left(\mathrm{SNR_{NL} - \mathrm{OSNR} - 3}\right)
    \quad \text{[dB]}.
\end{equation}
Formally, this rule is valid only when applied to a single span, or to the end-to-end system in 
the case of all identical, equally-spaced channels, and identical spans 
\cite{bosco:ofc2011,grellier:optExp2011}. 
Quite remarkably, it has been shown to work relatively well 
even under more general conditions, including multi-band
transmission \cite{jiang:ecoc2024}. 

Therefore, assuming that the total SNR and the transceiver SNR are known, 
the estimation of the nonlinear SNR (or the OSNR) can enable automatic power optimization, without requiring a detailed model of the network.

\subsection{NLI Estimation Methods}
Traditionally, NLI in a network can be estimated using two different approaches.
First, the OSNR can be measured at the receiver using an 
optical spectrum analyzer (OSA). Then, using \eqref{eq:snr},
together with a back-to-back calibration of $\mathrm{SNR_{TRX}}$,
the NLI can be computed.
This method is simple and effective, but it needs an OSA at every receiver
in the network, which can be impractical. Moreover, it requires
sufficient spacing between WDM channels to accurately measure the noise floor, which
reduces the overall spectral efficiency. \\
Second, it can be estimated using an NLI model within a digital twin of the optical network.
Modern models \cite{poggiolini:jlt2026,gan:jlt2026} allow accurate estimation of NLI
in real time thanks to closed-form expressions, 
but their accuracy depends on the knowledge of the link parameters. 

DSP-based NLI estimation methods can be classified according to different criteria. 
One possibility is to distinguish between time-domain methods, 
which exploit the statistical properties and temporal correlations of NLI, 
and frequency-domain methods, which analyze its spectral characteristics or introduce controlled spectral
perturbations \cite{pilori:sppcom2025,pilori:ecoc2026invited}. 
However, some techniques combine features from both domains. 
Alternatively, the methods can be classified according to the estimation 
approach, distinguishing between model-based and statistical methods, and 
machine learning (ML)-based methods. These two classifications are 
complementary, as both ML and non-ML techniques can exploit the same NLI 
signatures. In the following, we adopt the second classification.

Both categories exploit differences between ASE and
Kerr-induced NLI. In particular, while ASE is very well modeled by an
AWGN source, NLI originates from the nonlinear interaction
of the transmitted WDM channels \cite{dar:opex2013}.
For instance, a portion of NLI is in the form of nonlinear phase noise (NLPN) \cite{dar:jlt2016}, with
different statistical properties compared to laser phase noise. Moreover, NLI exhibits specific
temporal correlations, spectral characteristics and a specific dependence
on the launch power. Consequently, these algorithms exploit
one or more of these properties to obtain an estimate of NLI power.

To aid comparison, 
Tables~\ref{tab:nli_methods_model} and~\ref{tab:nli_methods_ml} summarize the main principles, 
inputs and limitations of the reviewed methods. 
These descriptions are 
intentionally high-level and do not include all implementation details or 
validation scenarios. In particular, training and calibration data refer to the preparation of the estimator, 
while side information refers to its operation.

\subsection{Model-based and Statistical Estimation Methods}
\begin{table*}[!t]
\centering
\caption{Model-Based and Statistical (non-ML) NLI Estimation Methods}
\label{tab:nli_methods_model}
\footnotesize
\setlength{\tabcolsep}{4pt}
\renewcommand{\arraystretch}{1.2}
\begin{tabularx}{\textwidth}{@{}R{2.0cm}Y{1.0}Y{0.9}Y{0.9}Y{1.0}Y{1.2}@{}}
\toprule
\textbf{Ref.} & \textbf{Principle} & \textbf{Training or calibration data} & \textbf{Main input features} &
\textbf{Tx/link intervention} & \textbf{Side information and main limitations}\\
\midrule
Dong \cite{dong:opex2012} (2012) & Temporal correlation of NLI &
Launch-power and OSNR sweeps & Received symbols &
Launch-power and OSNR sweeps (calibration) &
Requires link length; calibration depends on WDM load\\
Di Rosa \cite{dirosa:opex2020} (2020) & Phase-noise nature of NLI & None & Received and transmitted symbols & None &
Does not separate ASE; simulation only\\
Lu \cite{lu:opex2020} (2020) & Non-flat NLI spectrum & Launch-power sweep & Received signal spectrum & Launch-power sweep (calibration) &
Recalibration when WDM load or filtering change; relies on GN model\\
Vaquero-Caballero \cite{vaquerocaballero:jlt2022} (2022) & Power scaling of NLI & None & Pre-FEC BER &
Power changes on spectral sub-bands & Requires transceiver noise and SNR margin; relies on GN model\\
Guo \cite{guo:ecoc2024} (2024) & Non-flat NLI spectrum & None & Noise spectrum of received symbols &
Launch-power sweep; interferer on/off & Blind separation only proposed; no NLI accuracy reported\\
L. Andrenacci \cite{andrenacci:jlt2025} (2025) & LPM (RP1 model) & None & Received symbols and
accumulated chromatic dispersion &
None & Estimation only of self-channel interference.\\
\bottomrule
\end{tabularx}
\end{table*}

Model-based and statistical methods do not require ML training, 
but some rely on calibration procedures.
The main model-based methods are summarized in Table~\ref{tab:nli_methods_model}.
In \cite{dong:opex2012}, the authors exploit the temporal correlation of amplitude noise between received 
symbols to correct an error vector magnitude (EVM)-based OSNR estimate for the contribution of NLI. However, this method requires calibration, which depends on the transmission distance and WDM loading.
In \cite{dirosa:opex2020}, the authors separate instead the NLPN contribution from Gaussian noise using 
statistical moments and EVM. 
However, it cannot separate ASE from the Gaussian part of NLI, and it has been demonstrated
only in simulations in ideal scenarios. In \cite{lu:opex2020,guo:ecoc2024}, 
the authors instead exploit the spectral characteristics of NLI, compared to ASE, 
which has a fully-flat spectrum.
However, \cite{lu:opex2020} requires a calibration procedure and 
it relies on the GN model, requiring some knowledge of the link parameters, while \cite{guo:ecoc2024} 
requires launch-power or interferer on/off changes.
In \cite{vaquerocaballero:jlt2022}, the authors perturb portions of the spectrum
of the transmitted signal. This operation is effective, but it requires modifications
of the transmitter, which may be difficult to implement within existing
standards and multi-source agreements (MSAs), 
and requires sufficient operating margin when applied to live traffic.

Finally, there is LPM-based NLI estimation \cite{andrenacci:jlt2025}. Since it is based
on a model of nonlinear propagation \cite{vannucci:jlt2002}, it is part of this first family of 
methods. Compared to the others, it does not require any training or calibration of the estimator,
and it computes directly the power of NLI using only information available at the receiver (received symbols
and accumulated chromatic dispersion). 
As will be shown later, the main limitation in WDM networks
is that it estimates only self-channel interference, which is just a part of the total NLI.

\subsection{Machine Learning-based Methods}
\begin{table*}[!t]
\centering
\caption{Machine-Learning-Based NLI Estimation Methods}
\label{tab:nli_methods_ml}
\footnotesize
\setlength{\tabcolsep}{4pt}
\renewcommand{\arraystretch}{1.2}
\begin{tabularx}{\textwidth}{@{}R{2.0cm}Y{1.0}Y{0.9}Y{0.9}Y{1.2}@{}}
\toprule
\textbf{Ref.} & \textbf{Principle} & \textbf{ML model and training data} & \textbf{Main input features} &
\textbf{Side information and main limitations}\\
\midrule
Vaquero-Caballero \cite{vaquerocaballero:jocn2018} (2018) & Temporal correlation and phase-noise nature of NLI &
NN; simulations & Received symbols & Requires number of channels and accumulated CD; simulation only\\
Lippiatt \cite{lippiatt:ofc2020} (2020) & Temporal correlation of NLI & NN; experiments & Estimated carrier phase &
Same-link training with intrusive labels; estimates launch power, not NLI\\
Shiner \cite{shiner:ofc2020} (2020) & Temporal and polarization correlation of NLI & NN; simulations & Received symbols &
Requires transceiver noise, link length and accumulated CD\\
Cho \cite{cho:opex22} (2022) & Non-Gaussian NLI distortion & CNN; experiments & Received symbols &
Same-link training with intrusive labels; tested only on 3-span links\\
Al-Nahhal \cite{alnahhal:jlt2024} (2024) & Statistics of the received signal & RNN and CNN; simulations & Received symbols (before phase recovery) &
Requires number of channels; simulation only\\
Boertjes \cite{boertjes:jlt2024} (2024) & Temporal correlation of NLI & Decision trees; analytical model & Received symbols &
Requires accumulated CD; neglects SPM; analytical model only\\
I.~Andrenacci \cite{andrenacci:jocn2024} (2024) & NLI--PDL interaction & Classifiers; analytical model & SNR time series &
Requires PDL; classifies operation below/above optimal launch power; no NLI power estimate \\
Bajaj \cite{bajaj:ondm2024} (2024) & Spectral and temporal correlation of NLI & NN; simulations &
Received signal spectrum and symbols & Requires number of channels and accumulated CD; simulation only\\
Girard-Jollet \cite{girardjollet:ofc2025} (2025) & Non-flat NLI spectrum & Decision trees; simulations & Noise spectrum of received symbols &
Requires transceiver noise; depends on carrier phase recovery settings\\
Bajaj \cite{bajaj:ecoc2026} (2026) & Spectral and temporal correlation of NLI & NN; simulations or experiments &
Received signal spectrum and symbols & Requires accumulated CD; same-link training with intrusive labels (experiments)\\
Girard-Jollet \cite{girardjollet:ecoc2026} (2026) & Non-flat NLI spectrum (SPM and XPM separated) & NN; experiments &
Noise spectrum of received symbols & Requires transceiver noise, symbol rate and accumulated CD; same-link training with intrusive labels\\
\bottomrule
\end{tabularx}

\vspace{2pt}
\parbox{\textwidth}{\footnotesize
CNN: convolutional NN; NN: neural network; RNN: recurrent NN. Intrusive labels: training targets obtained from launch-power sweeps or noise measurements requiring changes to the WDM channels.}
\end{table*}
In contrast with the previous methods, these techniques leverage ML to replace analytical models with
trained ones. Several ML-based methods have been proposed in the literature, and the main ones are
summarized in Table~\ref{tab:nli_methods_ml}.
They can be broadly grouped according to the signatures used as input features.

The first group uses the temporal correlation of NLI, leveraging the received symbols or the
estimated carrier phase \cite{vaquerocaballero:jocn2018,lippiatt:ofc2020,shiner:ofc2020,boertjes:jlt2024}.
A second group uses the non-Gaussian statistics of the noise affecting the received constellations
\cite{cho:opex22,alnahhal:jlt2024}. A third group uses the non-flat spectrum of the noise
\cite{girardjollet:ofc2025,bajaj:ondm2024,girardjollet:ecoc2026,bajaj:ecoc2026}, similarly
to some non-ML methods \cite{lu:opex2020,guo:ecoc2024}. Finally, \cite{andrenacci:jocn2024}
exploits the interaction between NLI and PDL.

In general, ML-based methods can provide accurate NLI estimates without requiring an explicit analytical model. 
However, the reliance on training raises questions about the generalizability of the results.
In most cases, training was performed on relatively small datasets, and testing was limited to
specific scenarios. Therefore, their performance under arbitrary network conditions remains an open question.

\section{Longitudinal Power Monitoring}
Before discussing LPM-based NLI estimation, which is the main focus
of this paper, this section briefly introduces the history and
basic principles of LPM.

\subsection{Basic Principles}
LPM estimates the channel power evolution by exploiting the well-known
non-commutativity between chromatic dispersion and the Kerr nonlinear
effect \cite{agrawal:book2019}, which is the same principle underlying
the split-step Fourier method (SSFM) \cite{pilori:icton2017}
and digital backpropagation \cite{ip:jlt2008}.

In fact, some early demonstrations of LPM \cite{sasai:jlt2022} were based on a full time-domain
simulation of the transmitted channel (e.g., using the SSFM), and tried to find the link's
power profile that best approximates the received signal. 
This channel model allowed consideration of not only dispersion and Kerr effect, but also other effects,
such as filtering by wavelength-selective switches (WSSs). However,
the overall computational complexity was unsuitable for a real-time implementation in the transceivers.

Most LPM demonstrations focused on two main techniques, which enable the estimation of a power profile 
with a closed-form expression, without relying on expensive optimizations: Correlation-based
Method (CM) \cite{tanimura:ecoc2019pdp,tanimura:jlt2020} and LLS \cite{sasai:jlt2024}.
Both methods are based on a first-order regular perturbation model \cite{vannucci:jlt2002}
of nonlinear propagation, which is known to be valid for many practical cases in optical 
communications. 

It was later shown that the two methods are closely related, as an LLS solution can be obtained by 
deconvolving a CM profile \cite{sasai:jlt2023}.
For the remainder of this work, we will focus on the LLS-LPM algorithm.

\subsection{Linear Least-Squares LPM}
\begin{figure}[ht]
\centering
\begin{tikzpicture}[scale=0.9, transform shape, node distance=4mm and 4mm]
  \node[input, text width=31mm] (rx) {Received signal\\$A[L,n]$};
  \node[input, right=of rx, text width=31mm] (ref) {Reference signal\\$A_{\mathrm{ref}}[0,n]$};

  \node[process, below=8mm of rx, text width=31mm] (pert)
    {Extract nonlinear term\\$A_1=A[L,n]-A_0[L,n]$};
  \node[process, below=8mm of ref, text width=31mm] (matrix)
    {Build perturbation\\matrix $\mathbf G$};

  \coordinate (topL) at ($(pert.north)+(0,3mm)$);
  \coordinate (topR) at ($(matrix.north)+(0,3mm)$);
  \draw[arrow] (rx.south) -- (topL) -- (pert.north);
  \draw[arrow] (ref.south) -- (topR) -- (matrix.north);
  \draw[arrow] (ref.south) -- ++(0,-3.7mm)
       -| ([xshift=8mm]pert.north);

  \node[process, below=8mm of $(pert.south)!0.5!(matrix.south)$,
    anchor=north, text width=70mm] (lls)
    {Linear Least-Squares\\[-1pt]
     $\displaystyle\hat{\boldsymbol\gamma}'=
      [\operatorname{Re}(\mathbf G^\dagger\mathbf G)]^{-1}
       \operatorname{Re}(\mathbf G^\dagger A_1)$};
  \draw[arrow] (pert.south) -- ++(0,-3mm) -| ([xshift=-19mm]lls.north);
  \draw[arrow] (matrix.south) -- ++(0,-3mm) -| ([xshift=19mm]lls.north);

  \node[output, below=of lls, text width=70mm] (profile)
    {Normalized power profile $\hat\gamma'(z)$};
  \draw[arrow] (lls) -- (profile);
\end{tikzpicture}
\caption{Simplified LLS-LPM \cite{sasai:jlt2024} 
  processing chain. The nonlinear term $A_1$
 and the perturbation matrix $\boldsymbol{G}$ are computed separately from the
 received and reference (transmitted) signals.}
\label{fig:lpm_lls}
\end{figure}
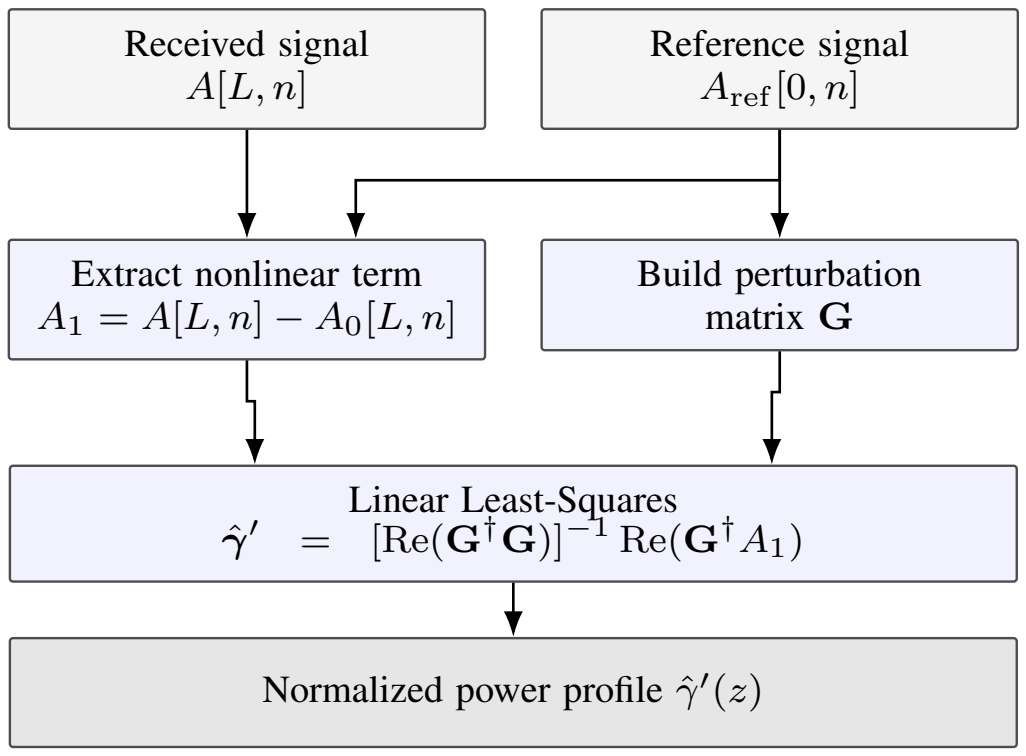
A high-level description of LLS-LPM is shown in Fig.~\ref{fig:lpm_lls}. As input, it requires
a few quantities, commonly available at the DSP of a coherent receiver.
First, it requires the received constellation after DSP (but before decoding), at one sample per symbol.
Then, it requires the total accumulated chromatic dispersion, which is routinely estimated by coherent DSP
\cite{sui:jlt2013}. It also requires the transmitted symbol sequence corresponding
to the received constellation. This can be obtained either using training sequences, or, more
simply, by taking hard decisions on the received constellation \cite{chang:ecoc2023,andrenacci:ofc2025,chang:ofc2025}. Decision
errors induce a deterministic power shift, which can be
compensated for, at the expense of slightly higher estimation noise.

From these data, two elements are built: the vector $\boldsymbol{A}_1$,
which is an estimate of the residual nonlinear perturbation at the receiver, and the perturbation
matrix $\boldsymbol{G}$.
Each column of $\boldsymbol{G}$ represents the effect of chromatic
dispersion applied from the beginning of the link to an intermediate
point, followed by a nonlinear operator, and then additional dispersion
up to the end of the link \cite{sasai:jlt2024}.
This matrix could be built assuming a regular grid in terms of distance
\cite{sasai:jlt2024}, but this requires knowledge of the 
dispersion map of the link. Alternatively, it can also be built \cite{pilori:ptl2025,pilori:jlt2026}
with a regular grid in terms of accumulated dispersion. In this case, 
no knowledge of the link length or the dispersion of the individual fiber spans is needed.
Afterwards, the power profile $\boldsymbol{\gamma}'$ is computed
from these two quantities.
Two main comments can be made on this operation.

First, this operation is memory intensive, especially for long links with small steps.
While this is not a major issue for offline processing on modern workstations, 
it is impractical for embedded applications. Therefore, there is growing research interest
\cite{sasai:ofc2026,chang:ecoc2026} in lower-complexity LPM implementations. However, for the purpose
of this work, implementation complexity will be neglected, and the full
LLS-LPM algorithm will be considered.

Second, the output of LPM is a ``normalized'' power profile $\boldsymbol{\gamma}'$, 
expressed as a function of accumulated chromatic dispersion rather than (physical) distance. 
To obtain the actual power profile, i.e., the absolute
optical power as a function of distance, some link parameters must be known. In particular, the nonlinear
coefficient $\gamma$ and dispersion parameter $\beta_2$ for each fiber span are required. 
The power profile can be corrected accordingly \cite{takahashi:oecc2025,pilori:jlt2026}. The advantage of this approach
is that the correction is a
simple scaling operation, which can be applied offline
after running the LLS-LPM algorithm. 
The only constraint is that the accumulated dispersion must be
a strictly monotonic function of the distance, i.e., no dispersion compensation must be
introduced in the link \cite{sasai:jlt2023,pilori:ptl2025}.

\section{NLI Estimation using LPM}

\subsection{Principles}
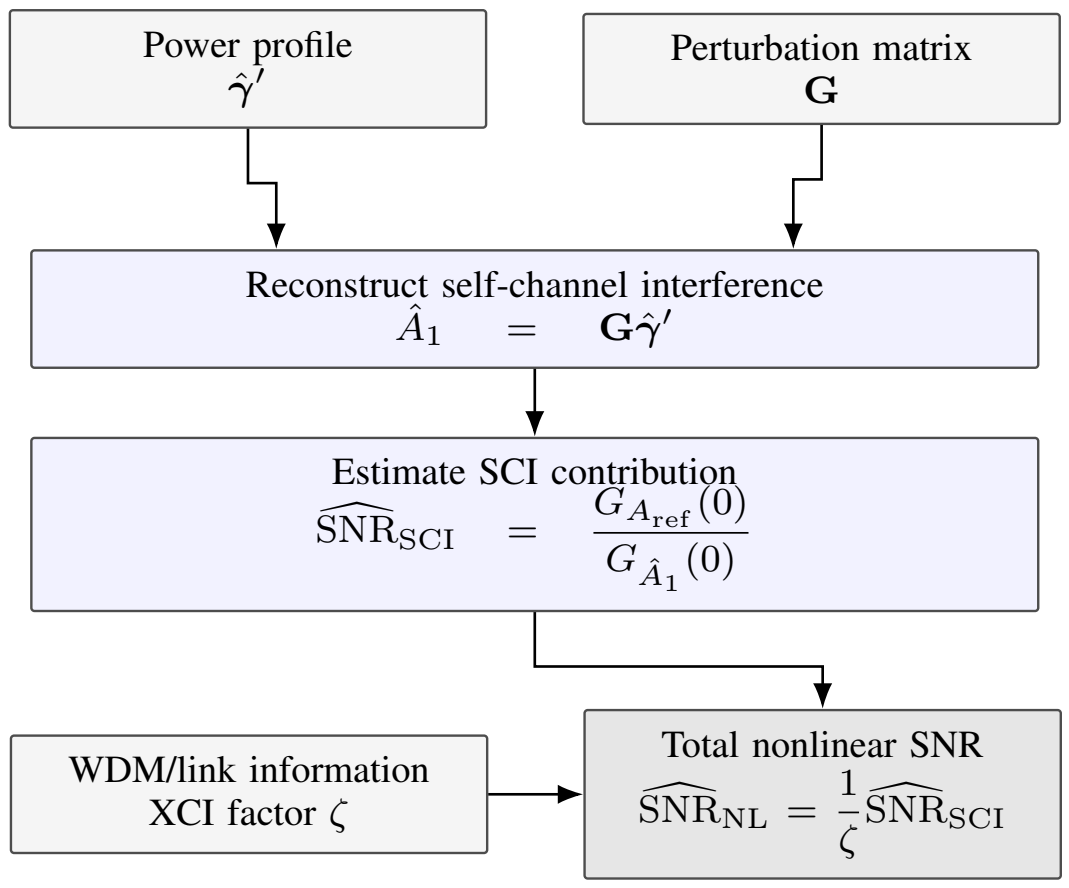
\begin{figure}[ht]
\centering
\begin{tikzpicture}[
    scale=0.9, transform shape,
    node distance=5mm and 7mm
]
  \node[input, text width=31mm] (profile)
    {Power profile\\$\hat{\boldsymbol\gamma}'$};

  \node[input, right=7mm of profile, text width=31mm] (matrix)
    {Perturbation matrix\\$\mathbf G$};

  \node[process,
    below=9mm of $(profile.south)!0.5!(matrix.south)$,
    anchor=north, text width=70mm] (recon)
    {Reconstruct self-channel interference\\
     $\hat A_1=\mathbf G\hat{\boldsymbol\gamma}'$};

  \draw[arrow] (profile.south) -- ++(0,-4mm)
    -| ([xshift=-19mm]recon.north);

  \draw[arrow] (matrix.south) -- ++(0,-4mm)
    -| ([xshift=19mm]recon.north);

  \node[process, below=5mm of recon, text width=70mm] (sci)
    {Estimate SCI contribution\\
     $\displaystyle\widehat{\mathrm{SNR}}_{\mathrm{SCI}}=
      \frac{G_{A_{\mathrm{ref}}}(0)}{G_{\hat A_1}(0)}$};

  \draw[arrow] (recon.south) -- (sci.north);

  \node[input, below=9mm of sci,
    xshift=-21mm, text width=31mm] (zeta)
    {WDM/link information\\XCI factor $\zeta$};

  \node[output, right=7mm of zeta,
    text width=31mm] (nli)
    {Total nonlinear SNR\\[1mm]
     $\displaystyle\widehat{\mathrm{SNR}}_{\mathrm{NL}}=
      \frac{1}{\zeta}\widehat{\mathrm{SNR}}_{\mathrm{SCI}}$};

  \draw[arrow] (sci.south) -- ++(0,-4mm)
    -| (nli.north);

  \draw[arrow] (zeta.east) -- (nli.west);

\end{tikzpicture}
\caption{NLI estimation using LPM \cite{andrenacci:jlt2025}.}
\label{fig:lpm_nli}
\end{figure}

The basic principle of LPM-based NLI estimation 
follows directly from the first-order regular perturbation (RP1)
model \cite{vannucci:jlt2002,serena:jlt2013}, 
which also forms the basis of LLS-LPM \cite{sasai:jlt2024}.
According to this model, the NLI at the receiver
can be reconstructed from the perturbation matrix
$\boldsymbol{G}$ and the normalized power profile
$\hat{\boldsymbol{\gamma}}'$ \cite{andrenacci:jlt2025}
\begin{equation}
\hat{\boldsymbol{A}}_1 =
\boldsymbol{G}\hat{\boldsymbol{\gamma}}'\,.
\end{equation}
A high-level schematic of this operation is shown in Fig.~\ref{fig:lpm_nli}.
This operation reconstructs the nonlinear perturbation from the LLS-estimated power profile, rather than directly using the residual perturbation extracted from the received signal.

Assuming that the NLI spectrum is approximately flat over the signal bandwidth \cite{poggiolini:jlt2012},
the nonlinear SNR can be computed from the power spectral
densities (PSDs) of the estimated NLI and the reference
(i.e., transmitted) signal
\begin{equation}
\widehat{\mathrm{SNR}}_\textup{SCI} =
\frac{G_{A_\textup{ref}}(0)}{G_{\hat{A}_1}(0)}\,.
\label{eq:snrnlest}
\end{equation}
Note that these expressions only require the normalized power
profile, without the correction described in the previous
section. Consequently, no knowledge of the link dispersion
map or nonlinear coefficients is required.

However, this method estimates only a portion of the total
NLI \cite{andrenacci:jlt2025}, namely the self-channel
interference (SCI) \cite{poggiolini:jlt2012},
sometimes also called intra-channel NLI \cite{dar:opex2013}, or loosely self-phase modulation (SPM).
This is the NLI generated by the nonlinear
interaction of the channel of interest (COI) with itself.
In a WDM network, however, NLI also originates from
interactions with the other channels, generating
cross-channel interference (XCI) and multi-channel
interference (MCI) \cite{poggiolini:jlt2012,carena:opex2014}.
While MCI is generally negligible for long, 
dispersion-uncompensated links with high symbol rates \cite{poggiolini:jlt2012}, XCI remains a
relevant NLI source, even if its relative contribution
is expected to decrease with the industry trend toward higher symbol rates
\cite{che:ecoc2026}.

In \cite{andrenacci:jlt2025}, we proposed to correct this
estimation using a simple scaling factor $\zeta$ representing
the ratio between the total NLI and SCI powers, neglecting MCI
\begin{equation}
\zeta = 1+\frac{P_\textup{XCI}}{P_\textup{SCI}}\,.
\label{eq:zetadef}
\end{equation}
The advantage of this factor is that it can be directly
applied to the nonlinear SNR estimated using
\eqref{eq:snrnlest}, obtaining an estimate of the total
nonlinear SNR
\begin{equation}
\widehat{\mathrm{SNR}}_\textup{NL} =
\frac{1}{\zeta}\widehat{\mathrm{SNR}}_\textup{SCI}\,.
\end{equation}
The factor $\zeta$ can be computed in different ways.
In \cite{andrenacci:jlt2025}, we derived an approximate formula:
\begin{equation}
\log_{10}(\zeta) \approx
\frac{1}{4}\log_{10}(N_\textup{ch})
-0.0475\left[
\log_{10}\left(\frac{B_\mathrm{L}}{B_\mathrm{R}}\right)
\right]^2\,.
\label{eq:zetaapprox}
\end{equation}
In this equation, $N_\textup{ch}$ is the total number of WDM
channels, while $B_\mathrm{L}$ and $B_\mathrm{R}$ are,
respectively, the optical bandwidths to the left and right
of the COI.

In \cite{pilori:jlt2026}, we compared this approximation with 
a more complete NLI model \cite{poggiolini:jlt2026}, finding good agreement
in several experimental scenarios, including heterogeneous
links and different modulation formats. However, the
simplified expression relies on the assumptions of 
homogeneous link and WDM loading, without ISRS,
and it may introduce additional errors in more
complex scenarios.

The main limitation of using more accurate NLI models is
the requirement for detailed knowledge of the link
parameters. If these parameters are available, they can
also be used to build a digital twin of the network and
predict the NLI power without relying on receiver DSP.
Consequently, for the remainder of this paper, we will
assume limited knowledge of the network and use
\eqref{eq:zetaapprox} to compute $\zeta$.

\subsection{Experimental Setup}
\begin{figure}[ht]
\centering
\includegraphics[width=\linewidth]{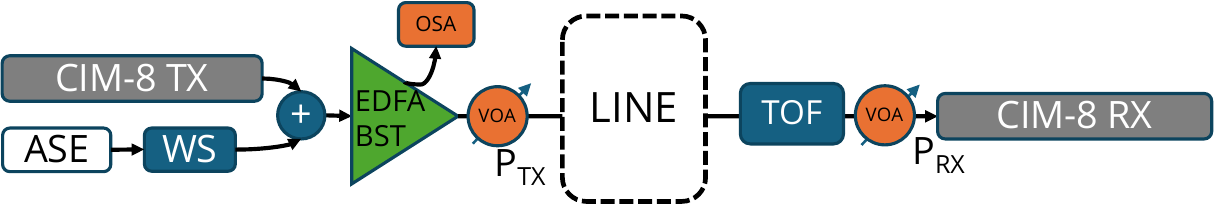}
\caption{Experimental setup for the results shown later (from \cite{pilori:jlt2026}).}
\label{fig:expsetup}
\end{figure}

To illustrate NLI estimation, we present some results obtained using the same experimental setup as 
in \cite{pilori:jlt2026}, summarized in Fig.~\ref{fig:expsetup}. The setup is based on the
transmission of 18 WDM channels at approximately 118 GBd.
The COI is generated using a commercial transceiver (CISCO CIM-8 equipped with 
Acacia's JANNU DSP), while the other WDM channels, spaced by 175~GHz, 
are emulated using ASE noise \cite{elson:opex2017,richter:ofc2018} 
spectrally shaped by a WaveShaper (WS). The COI is modulated using 
uniform 16-QAM for the short link and QPSK for the long link.

The WDM comb is then transmitted over two different links, one short and one long.
The short link consists of 5 spans of G.652 single-mode fiber (SMF), 
each approximately 85 km long, for a total link length of 431 km.
The long link, instead, consists of 8 spans of SMF, followed by 4 spans of G.654 pure-silica-core 
fiber (PSCF), and then another 8 spans of SMF.
The SMF spans are approximately 65 km long, while the PSCF spans are approximately 108 km long, for a 
total link length of 1483 km. In both links, each span is followed by an erbium-doped fiber amplifier 
(EDFA) to compensate for the span loss.

After transmission, the COI is selected using a tunable optical filter (TOF) and detected by the 
real-time DSP of the CIM-8 transceiver.
From the transceiver telemetry data, we captured the average bit error ratio (BER), the accumulated 
CD, and several constellation blocks, each $12\,288$ symbols long.
Each block was then used as input to the LLS-LPM algorithm, followed by NLI estimation.
The reference signal was generated by taking hard decisions on the received constellation.
The resulting estimation bias was corrected using the symbol error 
rate \cite{andrenacci:ofc2025,chang:ofc2025}, computed from the BER.

\subsection{Power Profiles}\label{sec:powprof}
\begin{figure*}[ht]
\centering
\includegraphics[width=0.9\linewidth]{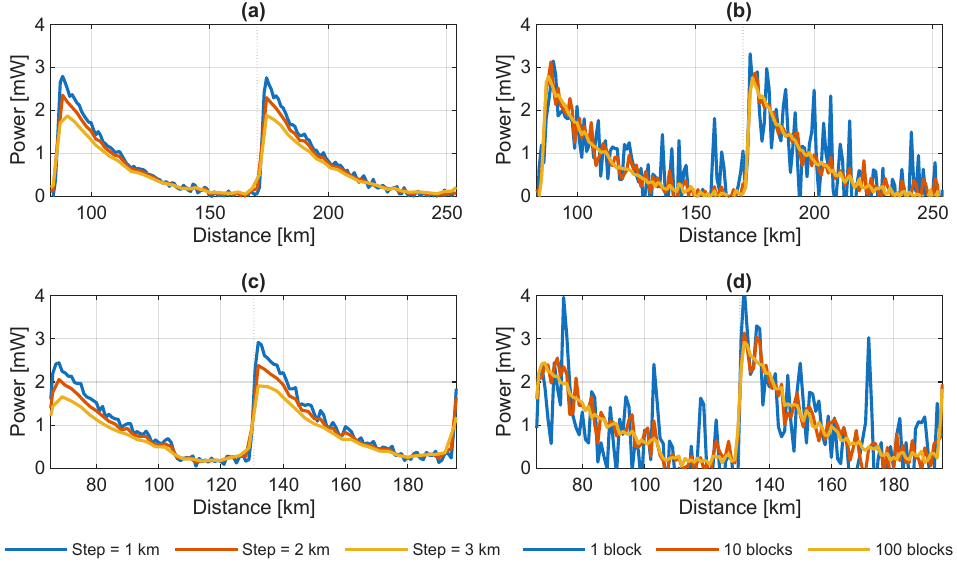}
\caption{Portions of the estimated power profiles for the
(a,b) 431-km and (c,d) 1483-km links, showing the impact
of (a,c) different spatial step sizes and (b,d) averaging
over different constellation blocks.
For (a,c), the results are averaged over 100 constellation blocks. For (b,d), the spatial step is fixed at 1 km.}
\label{fig:profiles}
\end{figure*}
The primary output of LLS-LPM is the power profile.
Fig.~\ref{fig:profiles} shows portions of the power profiles for the short (a,b) and long (c,d) 
links, corresponding to the second and third spans.
The profiles were computed on a uniform spatial grid,
using the link's dispersion map to build the
$\boldsymbol{G}$ matrix. The absolute power was then
computed using knowledge of the nonlinear coefficient
$\gamma$ of each span.
Although this information is not required for NLI
estimation, it is generally needed to obtain the
actual optical power as a function of distance,
which is the main purpose of LPM.
Moreover, using a uniform spatial grid is consistent
with previous experimental demonstrations and
allows a direct comparison with the current literature.

The plots illustrate the impact of the spatial step size and the number of averaged constellation blocks.
The total launch power is approximately 18.5 dBm (6 dBm per channel) for both links.

Figs.~\ref{fig:profiles}(a) and (c) show the impact of the LLS-LPM step size, ranging from 1 to 3 km. 
As predicted by theory \cite{sasai:jlt2025}, a larger step size increases robustness and reduces estimation noise, at the expense of spatial resolution.
This is particularly clear when comparing the estimated power at the beginning of each span.
An optimal step size can be computed using the theoretical derivation in \cite{sasai:jlt2025}. 
For instance, in \cite{kaneko:ofc2026}, the authors derived an optimal step size for detecting a 1-dB 
anomaly.
However, as will be shown later, the step size has a different impact on NLI power estimation.

Figs.~\ref{fig:profiles}(b) and (d) show the impact of averaging over different numbers of 
constellation blocks (each $12\,288$ symbols long). The step size was kept fixed at 1 km.
With a single block, the profile is very noisy, although the power evolution can still be
identified.
Increasing the number of blocks results in a smoother profile.
Note that this improvement is possible because each constellation block contains a different, random 
transmitted symbol sequence \cite{sasai:ecoc2024}.

\subsection{NLI Estimation Accuracy}
\begin{figure}[ht]
\centering
\includegraphics[width=0.9\linewidth]{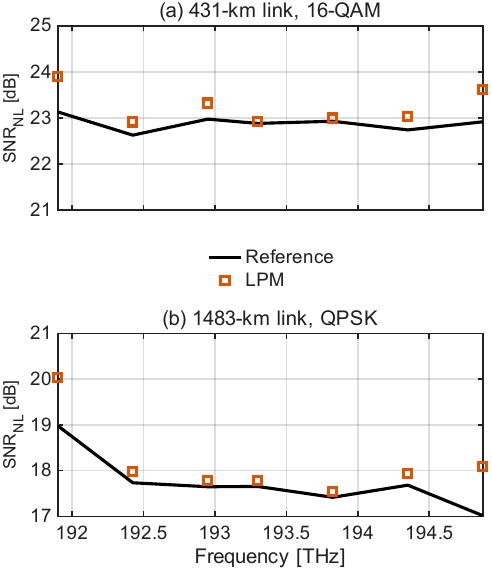}
\caption{Accuracy of NLI estimation using LPM, for the short (a) and long (b) links. Markers: LPM estimates; solid lines: reference obtained from the measured BER and the OSNR measured with an OSA. Adapted from 
\cite{pilori:jlt2026}.}
\label{fig:nliestimation}
\end{figure}
Fig.~\ref{fig:nliestimation}
summarizes results reported in \cite{pilori:jlt2026} for NLI estimation accuracy.
The results are shown for the short (a) and long (b) links. 
The spatial step is fixed at 1 km, 
and the results are averaged over 100 constellation blocks. The correction factor $\zeta$ is computed using
\eqref{eq:zetaapprox}. 
The reference nonlinear SNR is obtained from the measured BER, after removing the transceiver noise using a back-to-back calibration and subtracting the ASE contribution in linear units.
For the short link, LPM generally overestimates the
nonlinear SNR, with deviations up to approximately 0.8~dB.
For the long link, instead, the agreement is closer for
the internal channels, while larger deviations are observed
at the edges of the measured band.

\subsection{Impact of Step Size and Block Averaging}
\begin{figure}[ht]
\centering
\includegraphics[width=0.85\linewidth]{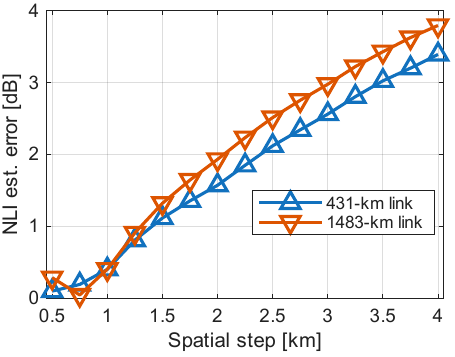}
\caption{NLI estimation error as a function of the spatial
step size for the 431-km and 1483-km links.
The error is computed from the ratio between the
averaged normalized NLI powers obtained using LPM
and the experimental reference, using 100
constellation blocks.}
\label{fig:lpmnumericalstep}
\end{figure}
\begin{figure}[ht]
\centering
\includegraphics[width=0.9\linewidth]{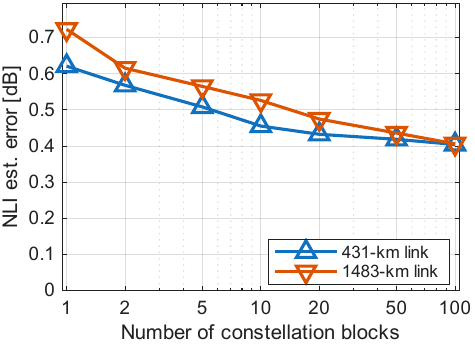}
\caption{Maximum NLI estimation error as a function of
the number of averaged constellation blocks for the
431-km and 1483-km links. The spatial step is fixed
at 1 km, and the maximum error is computed over
the different averaging windows within the dataset
of 100 blocks.}
\label{fig:lpmblocknumber}
\end{figure}
In Sec.~\ref{sec:powprof},
we discussed the impact of the spatial step size on the estimated power profiles, which
has been extensively studied \cite{sasai:jlt2025,kaneko:ofc2026}. 
Figs.~\ref{fig:lpmnumericalstep} and \ref{fig:lpmblocknumber} show the impact of the spatial step size and the number of averaged constellation blocks, respectively, on the NLI estimation error.
To obtain a consistent comparison, we first compute the normalized NLI power, 
$1/\mathrm{SNR}_{\mathrm{NL}}$, and average it in linear units over all the
evaluated wavelengths, separately for the corrected LPM estimates and the experimental reference. 
The estimation error is then computed as the absolute value, in dB, of the ratio between these two averages.
For each block count, we report the maximum error over all contiguous windows of that length.

Fig.~\ref{fig:lpmnumericalstep} shows this error for different spatial steps. 
Since most of the NLI is generated near the beginning of each span, a small spatial step is required to 
accurately reconstruct the power profile immediately after the amplifiers, as seen in 
Fig.~\ref{fig:profiles}. Consequently, the spatial step has a strong impact on the NLI
estimation error, with errors below about 0.5 dB for steps up to 1 km, rapidly increasing for larger steps. 
For the long link, the error slightly increases at the smallest step size (500 m), 
possibly due to increased estimation noise when operating far from the ``optimal'' step size 
\cite{kaneko:ofc2026} for 118 GBd signals.

In contrast, the number of constellation blocks has a limited impact on NLI
estimation accuracy, since averaging mainly improves the accuracy of the power profile in low-power regions.
Fig.~\ref{fig:lpmblocknumber} shows the maximum estimation error obtained by averaging
different numbers of blocks within the dataset of 100 blocks. The error decreases only
slightly as the number of blocks increases. 
Remarkably, even with a single block, the maximum NLI estimation error remains below 0.75 dB, despite the 
significant noise affecting the estimated power profiles.

\section{Discussion}

\subsection{LPM-based NLI Estimation}
In the previous section, we investigated the impact of the spatial
step size and the number of averaged constellation blocks
on NLI estimation accuracy.
The results show that the requirements for NLI estimation
are different from those for power profile reconstruction.
In fact, most of the NLI is generated in the high-power
regions, near the beginning of each span.
Consequently, an accurate reconstruction of these regions
is more important than obtaining a smooth power profile
over the entire link.
This explains why the spatial step size has a significant
impact on NLI estimation, while averaging additional
constellation blocks provides a relatively small improvement.

This result is particularly interesting for practical
implementations, since it suggests that accurate NLI
estimation does not necessarily require a highly accurate
power profile.

\subsection{Practical Limitations}
The main limitation of LPM-based NLI estimation is that
it estimates only SCI, rather than the total NLI.
As discussed previously, this limitation can be addressed
using a correction factor derived from analytical NLI models.
However, the accuracy of this correction depends on the
assumptions of the model and on the available knowledge
of the network.
In particular, heterogeneous WDM loading and unequal
channel powers may introduce additional estimation errors
\cite{pilori:jlt2026}.

Another limitation is the reduced estimation accuracy when
NLI is weak compared to the other noise contributions.
At the optimal launch power \eqref{eq:distFromOpt}, NLI power is not the
dominant noise component, and it becomes even weaker
at lower launch powers.
In this regime, the stronger relative contribution of ASE
and transceiver noise
reduces the accuracy of the LPM estimation
\cite{andrenacci:jlt2025,pilori:jlt2026}.
However, this limitation is not specific to LPM, but is
common to other receiver-based NLI estimation methods.

Finally, obtaining a reliable experimental benchmark
remains challenging.
As discussed in Sec.~\ref{sec:nliest}, there is no
standard and accurate method to directly measure NLI power.
Even the OSA-based approach used as a reference in this
work is subject to measurement uncertainties and requires
sufficient spacing between WDM channels.
Since the reference NLI is computed by removing
different noise contributions, relatively small
measurement errors may result in significant uncertainties,
especially when NLI is weak compared to the other noise sources.

\subsection{Network Applications}
The main application of NLI estimation is the optimization
of the launch power.
As discussed in Sec.~\ref{sec:nliest}, the knowledge of
the nonlinear SNR and OSNR allows estimating the optimal
launch power with little knowledge of the network's parameters.

However, end-to-end NLI estimation provides only an
aggregate metric, which is not sufficient to independently
optimize the launch power in each span of a heterogeneous
link.
In this context, LPM offers an additional advantage:
the possibility of obtaining span-resolved performance metrics,
providing information on signal degradation
at different positions along the link.
Recent works have investigated this possibility
\cite{hahn:ptl2025,pilori:iccc2025,hahn:ecoc2026},
opening the way to automatic span-wise
power optimization.

Nevertheless, obtaining span-resolved estimates of the NLI contributions 
is significantly more challenging than end-to-end estimation. 
In fact, in a standard long-distance link, the NLI generated by each span 
is only a small fraction of the total accumulated NLI. 
In particular, the per-span nonlinear SNR can easily exceed the back-to-back transceiver SNR, which can be 
around 20~dB in modern commercial transceivers \cite{pfeifle:ofc2025}. 
Moreover, the per-span NLI contribution is obtained from the difference between two cumulative estimates, 
making it more sensitive to estimation errors. Consequently, the accuracy of span-resolved NLI estimation is 
expected to be lower than that of end-to-end estimation.

Beyond launch-power optimization, NLI estimates could
also be used as additional telemetry information for
network management.
For instance, separating ASE and NLI contributions
would help identify the origin of performance degradations,
while the availability of NLI estimates from several
channels could support network-wide optimization and
the validation of digital twins.

\subsection{Next Steps and Future Work}
Despite recent advances, several research challenges
remain open.

First, improving the estimation of the XCI contribution
is an important direction.
The correction factors discussed in this paper are based
on analytical models and require some knowledge of the
WDM configuration.
An interesting alternative could be to exploit the
SCI estimates obtained from multiple channels in the
same network.
For instance, the SCI measured at different wavelengths
could provide additional information on the nonlinear
characteristics of the link, potentially improving the
estimation of XCI without requiring complete knowledge
of the network parameters.
However, the relationship between SCI measurements and
the XCI affecting each channel would need to be
investigated, especially in heterogeneous networks.

A second research direction is the application of NLI
estimation to automatic network optimization.
Although the theoretical relationship between NLI and
optimal launch power is well established, practical
implementations based on receiver-side NLI estimates
are still relatively limited.
In particular, the combination of end-to-end and
span-wise NLI estimates could enable power optimization
in heterogeneous links, while accounting for different
fiber types, amplifier configurations and WDM loading.
The impact of estimation errors on the resulting
optimization performance also deserves further investigation.

Finally, an important challenge is reducing the
computational complexity of NLI estimation.
Recent results \cite{sasai:ofc2026pdp,sasai:ecoc2026invited} have already
demonstrated the feasibility of implementing low-complexity LPM-based
NLI estimation directly within a coherent DSP.
However, further complexity reductions could be obtained
by designing algorithms specifically for NLI estimation,
rather than for the reconstruction of the entire power
profile.
In fact, the results presented here suggest that NLI
estimation is less sensitive to the noise affecting
the estimated profile in low-power regions.
This could also motivate approaches based on nonuniform
spatial discretization without explicitly
reconstructing the full power profile.
Such techniques could further reduce the computational
and memory requirements, making continuous NLI monitoring
more practical in future-generation coherent transceivers.

\section{Conclusion}
In this invited paper, we reviewed the main techniques for NLI estimation in coherent optical systems, 
with particular focus on LPM-based methods. 
We discussed their principles, advantages and limitations, 
and presented new experimental results on the impact of spatial resolution 
and block averaging on NLI estimation accuracy. 
The results show estimation errors below approximately 0.5 dB for spatial steps up to 1 km, 
and below 0.75 dB even when using a single constellation block. 
Remarkably, accurate NLI estimation does not necessarily require an equally accurate reconstruction 
of the power profile, suggesting opportunities for further complexity reduction. 
Despite the limitations associated with SCI-only estimation and the need for XCI correction, 
LPM-based NLI estimation represents a promising technique for receiver-side monitoring and optical power optimization.

\section*{Acknowledgment}
The author would like to express his gratitude to Lorenzo Andrenacci for his fundamental contributions 
to the early development of this research activity. 
Special thanks go to Gabriella Bosco and Ann Margareth
Rosa Brusin for their continuous support and many fruitful discussions.

The author also thanks Antonino Nespola and Stefano Straullu 
from LINKS Foundation for carrying out the experimental work presented in this paper.

This research activity was partially supported by CISCO through sponsored research agreements in 2024 and 2025.

\bibliographystyle{IEEEtran}
\bibliography{bibliography}

\end{document}